\input{aipcheck}

\documentclass[
    ,final            
  ]
  {aipproc}

\layoutstyle{8x11single}

\begin{document}

\title{Are Galaxy Clusters Suggesting an Accelerating Universe?}


\author{J. A. S. Lima}{
  address={Departamento de Astronomia, Universidade de S\~{a}o
Paulo \\ Rua do Mat\~ao, 1226 - 05508-900, S\~ao Paulo, SP, Brazil},email={limajas@astro.iag.usp.br}
}

\author{R. F. L. Holanda}{
    address={Departamento de Astronomia, Universidade de S\~{a}o
Paulo \\ Rua do Mat\~ao, 1226 - 05508-900, S\~ao Paulo, SP, Brazil},email={holanda@astro.iag.usp.br}
}

\author{J. V. Cunha}{
   address={Departamento de Astronomia, Universidade de S\~{a}o
Paulo \\ Rua do Mat\~ao, 1226 - 05508-900, S\~ao Paulo, SP, Brazil},email={cunhajv@astro.iag.usp.br}ss
}

\begin{abstract}
The present cosmic accelerating stage is discussed through a new kinematic method based on the Sunyaev-
Zel'dovich effect (SZE) and X-ray surface brightness data from galaxy clusters. By using the SZE/X-ray data from 38 galaxy
clusters in the redshift range $0.14 \leq z \leq 0.89 $ [Bonamente et al., Astrop. J. {\bf 647}, 25 (2006)] it is found that the present Universe
is accelerating and that the transition from an earlier decelerating to a late time accelerating regime is relatively recent. The
ability of the ongoing Planck satellite mission to obtain tighter constraints on the expansion history through SZE/X-ray angular
diameters is also discussed. Our results are fully independent on the validity of any metric gravity theory, the possible matter-
energy contents filling the Universe, as well as on the SNe Ia Hubble diagram from which the presenting accelerating stage
was inferred.\end{abstract}

\maketitle


\section{Introduction}

One decade ago, observations from distant type Ia supernovae lead to a landmark conclusion: the universal expansion is
speeding up and not slowing down as believed since the early days of observational cosmology \cite{riess98,perl99}. This phenomenon
is normally interpreted as a dynamic influence of some sort of dark energy whose main effect is to change the sign of
the universal decelerating parameter q(z)\cite{Padm03}. Another possibility is that the cosmic acceleration is a manifestation of
new gravitational physics (rather than dark energy) that involves a modification of the left hand side (geometric sector)
of the Einstein field equations. In this kind of theory the Friedmann equation is modified and a late time accelerating
stage of the Universe is obtained even for a Universe filled only with cold dark matter (CDM). At present, the space
parameter associated with the cosmic expansion is too degenerate, and as such, it is not possible to decode which
mechanism or dark energy component is operating in the cosmic dynamics \cite{Padm03,FR}.

SNe type Ia are not only the powerful standard candles available but still provides a unique direct access to the late
time accelerating stage of the Universe. Naturally, this a rather uncomfortable situation from the observational and
theoretical viewpoints even considering that ten years later, the main observational concerns about errors in SNe type
Ia measurements, like host galaxy extinction, intrinsic evolution, possible selection bias in the low redshift sample
seem to be under control \cite{Kowalski08}. A promising estimator fully independent of SNe type Ia and other calibrators of the
cosmic distance ladder is the angular diameter distance (DA(z)) from a given set of distant objects. It has also been
recognized that the combination of SZE  and X-ray surface brightness measurements may provide useful angular
diameters from galaxy clusters \cite{sunzel70,caval77,Birk99,SZEPapers2,Boname06,CML07}.

On the other hand, since the mechanism causing the acceleration is still unknown, it is interesting to investigate
the potentialities of SZE/X-ray technique from a more general viewpoint, that is, independent of the gravity theory
and the matter-energy contents filling the Universe. The better strategy available so far is to consider the same kind
of kinematic approach which has been successfully applied for determining the transition deceleration/acceleration in
the past by using SNe type Ia measurements \cite{TR02,Riess04,EM06,CL08,Cunha09}.

In this work, we employ a purely kinematic description of the universal expansion based on angular diameter
distances of clusters for two different expansions of the deceleration parameter. As we shall see, by using the
Bonamente et al. \cite{Boname06} sample we find that a kinematic analysis based uniquely on cluster data suggests that the Universe
undergone a dynamic phase transition (deceleration/acceleration) in a redshift $z \approx  0.3$. Further, it is also shown that
the Planck satellite mission data must provide very restrictive limits on the space parameter, thereby opening an
alternative route for accessing the expansion history of the Universe.
\section{Angular Diameter and Kinematic Approach}

 Let us now assume
that the Universe is spatially flat as motivated by inflation and
WMAP measurements \cite{Komat08}. In this case, the angular diameter
distance in the FRW  metric is defined by (in our units $c=1$),
\begin{eqnarray}\label{eq:dLq}
D_A &=& (1+z)^{-1}H^{-1}_{0}\int_0^z {du\over H(u)} = \frac{(1+z)^{-1}}{H_0} \nonumber \\
&& \,\, \int_0^z\, \exp{\left[-\int_0^u\, [1+q(u)]d\ln
(1+u)\right]}\, du,
\end{eqnarray}
where  $H(z)=\dot a/a$ is the Hubble parameter, and, $q(z)$, the
deceleration parameter, is defined by

\begin{eqnarray}\label{qz}
q(z)\equiv -\frac{a\ddot a}{\dot a^2} = \frac{d H^{-1}(z)}{ dt} -1.
\end{eqnarray}

In the framework of a flat FRW metric, Eq. (1) is an exact
expression for the angular diameter distance. As one may check, in
the case of a linear two-parameter expansion for $q(z)=q_0+z{q_1}$,
the above integral can  analytically be represented as

\begin{eqnarray}\label{eq:dAKin}
D_A(z) &=& \frac{(1+z)^{-1}}{H_0}e^{q_1}q_1{^{q_0-q_1}}
[\gamma{({q_1-q_0},(z+1){q_1})} \nonumber \\
&& \,\, - \gamma{({q_1-q_0},{q_1})}],
\end{eqnarray}
where ${q_0}$ and ${q_1}$ are the values of $q(z)$ and its redshift
derivative, $dq/dz$ evaluated at $z=0$ while $\gamma$ is an
incomplete gamma \cite{AbrSte72}. By using the above expressions we
may get information about $q_0$, $q_1$ and, therefore, about the
global behavior of $q(z)$. In principle,  a dynamic ``phase
transition" (from decelerating to accelerating) happens at
$q(z_t)=0$, or equivalently, $z_t=-q_0/q_1$. Another interesting
parametrization  is $q(z)=q_o + q_1 z/(1+z)$ \cite{CL08,Cunha09}. It
has the advantage to be well behaved at high redshift while the
linear approach diverges at the distant past. Now, the integral (1)
assumes the form:
\begin{eqnarray}\label{eq:dAKin2Parame}
D_A (z) &=& \frac{(1+z)^{-1}}{H_{0}}e^{q_{1}}{q_1^{-(q_0+q_1)}}
[\gamma(q_{1}+q_{0},q_1) \nonumber \\
& & \,\, - \gamma{({q_1+q_0},q_1/(1+z))}],
\end{eqnarray}
where  ${q_1}$ now is the parameter yielding the total correction in
the distant past ($z\gg 0, q(z)= q_0 + q_1$) and $\gamma$ is again
the incomplete gamma function.  Note that in this case the
transition redshift is defined by $z_t=-q_0/(q_0+q_1)$.

\begin{figure}[t]
\resizebox{.45\textwidth}{!}
{\includegraphics{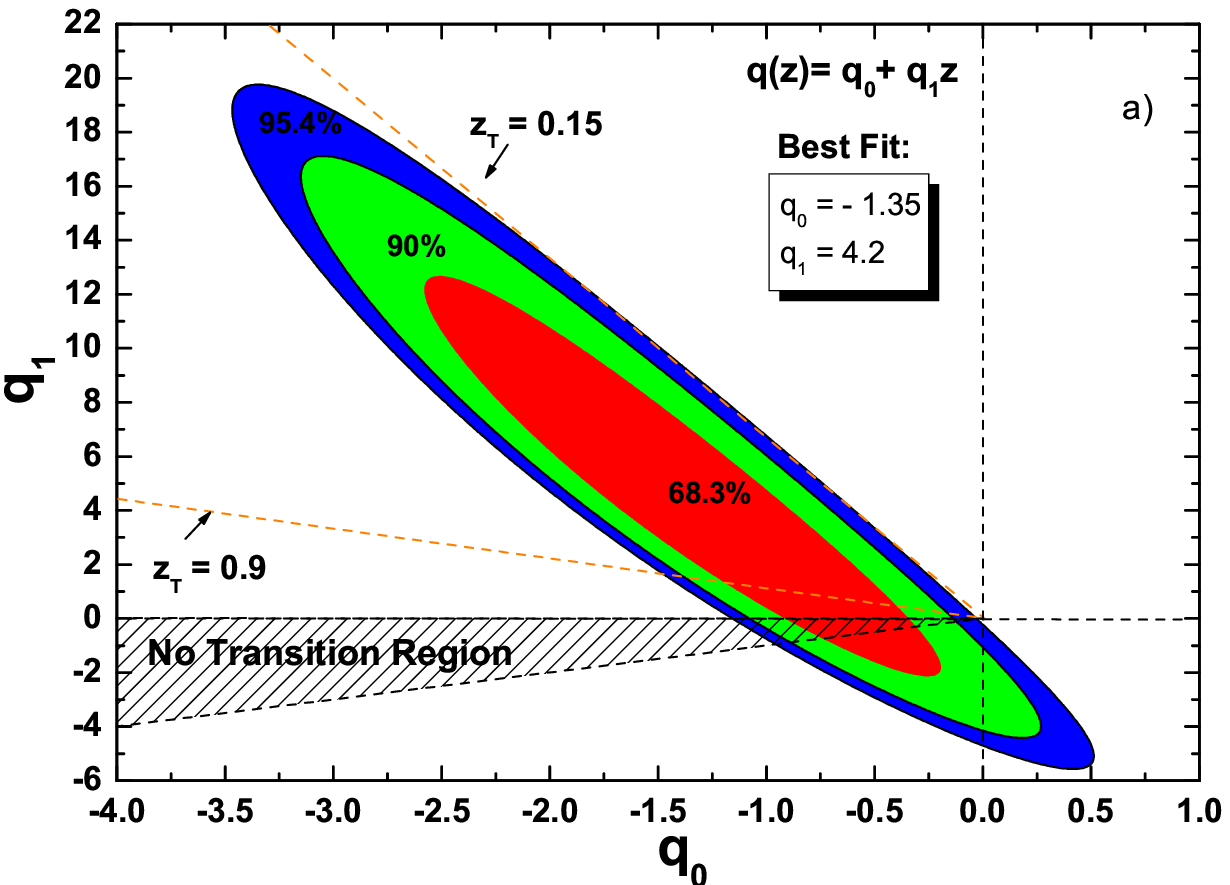}}
\resizebox{.451\textwidth}{!}
{\includegraphics{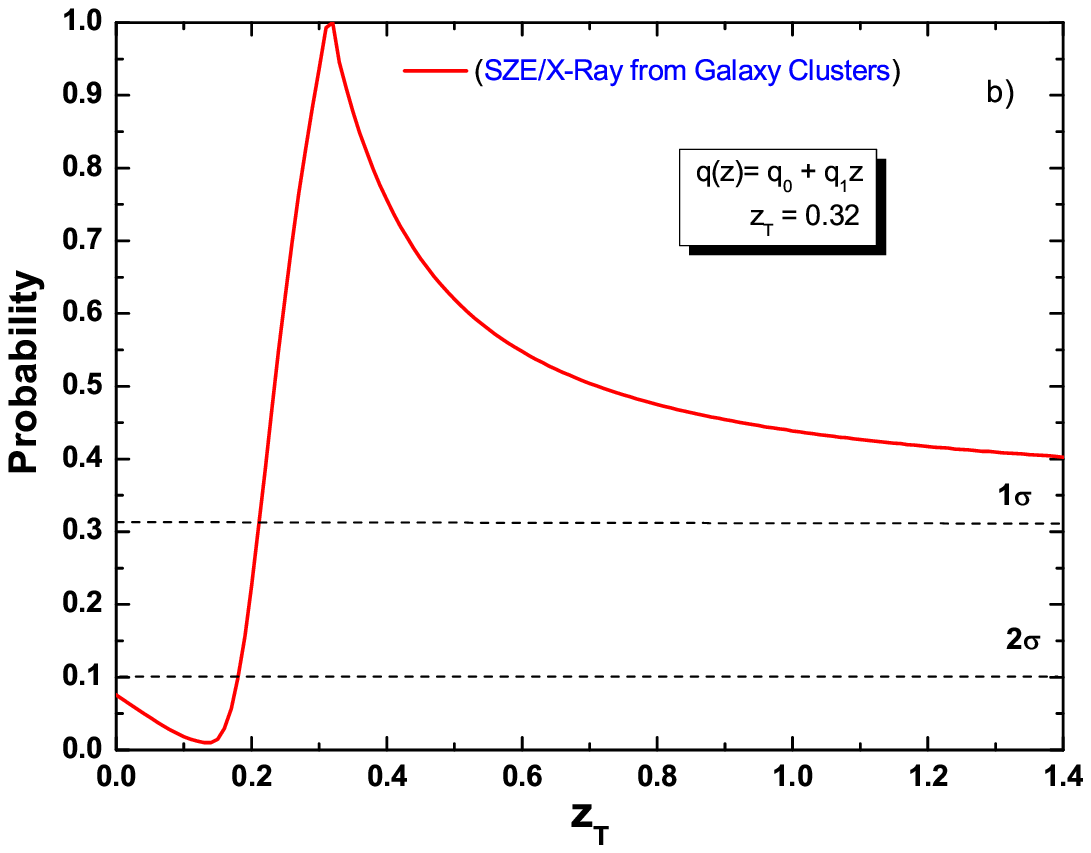}}
\caption{{\bf{a)}} Contours in the $q_o - q_1$ plane for 38 galaxy
clusters data \cite{Boname06} considering $q(z)=q_0 +q_1z$. The best
fit to the pair of free parameters is ($q_0,q_1) \equiv
(-1.35,4.2)$. For comparison we have shown the straight lines
denoting the transitions redshifts for two different flat $\Lambda$CDM
models: $z_t=0.9$ for $\Omega_{\Lambda}=0.8$ and $z_t=0.15$ for
$\Omega_{\Lambda}=0.43$. {\bf{b)}} Probability for the
transition redshift. The best fit is $z_t = 0.32$.}\label{fig1}
\end{figure}
\section{SZE/X-ray Technique and Constraints} 

The SZE is a small
distortion on the Cosmic Microwave Background (CMB) spectrum
provoked by the inverse Compton scattering of the CMB photons
passing through a population of hot electrons. This distortion is characterized by a low frequency ($< 218$ GHz) decrement and higher frequency ($> 218$ GHz) increment in the CMB intensity.  The measured
temperature decrement $\Delta T_{\rm SZ}$ in the Raylegh- Jeans frequencies of the CMB is given by:
\begin{equation}
\label{eq:1}
\frac{\Delta T_{\rm SZ}}{T_{\rm CMB}} = f(\nu, T_{\rm e}) \frac{ \sigma_{\rm T} k_{\rm B} }{m_{\rm e}}
\int _{\rm l.o.s.}n_e(r) T_{\rm e}(r) dl \
\end{equation}
where  $T_{\rm e}$ is the temperature of the intra cluster medium (ICM), $n_e$ is the electron density,  $k_{\rm B}$ the
Boltzmann constant, $T_{\rm CMB} =2.728^{\circ}$K is the temperature of the
CMB, $\sigma_{\rm T}$ the Thompson cross section, $m_{\rm e}$ the
electron mass and $f(\nu, T_{\rm
e})$ accounts for frequency shift and relativistic corrections ($c=1$ in our units). The integral is evaluated along the line of sight (l.o.s.).

The hot gas in galaxy clusters is also responsible for X-ray emission due to bremsstrahlung and line radiation resulting 
from electron-ion collisions. The X-Ray surface brightness $S_X$ is proportional to
the integral along the line of sight of the square of the 
electron density:
\begin{equation}
S_X = \frac{D^{2}_{A}}{4 \pi D^{2}_{L}} \int _{\rm l.o.s.} n_e^2(r)
\Lambda_{eH} dl
\label{eq:2}
\end{equation}
where $\Lambda_{eH}$ is the X-Ray cooling function of the ICM in the
cluster rest frame and $D_{L}$ is the luminosity distance from the galaxy cluster.
Observing the decrement of temperature of the CMB in the direction of galaxies clusters and also considering the X-rays observations, which are sensible to a different combination of the cluster electron density $n_e$ and the temperature $T_e$, it is possible to break the degeneracy between concentration and temperature and to calculate  the distance of angular diameter, yielding
\begin{equation}
D_{A}\propto \frac{(\Delta T_{SZ})^{2}\Lambda_{eH}}{S_{X}T^{2}_{e}}\frac{1}{\theta_{c}}
\end{equation}
where $\theta_{c}$ refers to a characteristic scale of the cluster along the l.o.s., whose exact meaning depends on the density model adopted. This technique for measuring distances is completely independent of other techniques and it can be used to measure distances at high redshifts directly.

Let us now consider the 38 measurements of angular diameter
distances  from galaxy clusters as obtained through SZE/X-ray method
by Bonamente and coworkers \cite{Boname06}. The cluster plasma and dark matter distributions were analyzed using a hydrostatic equilibrium model (assuming  spherical symmetry) that accounts for radial variations in density, temperature and abundance. 

In our analysis we use a
maximum likelihood determined by a $\chi^{2}$ statistics

\begin{equation}
\chi^2(z|\mathbf{p}) = \sum_i { ({{D}}_A(z_i; \mathbf{p})-
{{D}}_{Ao,i})^2 \over \sigma_{{{D}}_{Ao,i}}^2 +
\sigma_{stat}^{2}}, \label{chi2}
\end{equation}
where ${{D}}_{Ao,i}$ is the observational angular diameter
distance, $\sigma_{{{D}}_{Ao,i}}$ is the uncertainty in the
individual distance, $\sigma_{stat}$ is the contribution of the
statistical errors added in quadrature ($\approx 20$\%) and the
complete set of parameters is given by $\mathbf{p} \equiv (H_{0},
q_{0},q_{1})$. The common errors
are: SZE point sources $\pm 8$\%, X-ray background$\pm 2$\%, 
Galactic N$_{H}$ $\leq \pm 1\%$, $\pm 15$\% for cluster asphericity, $\pm 8$\% kinetic SZ and for CMB anisotropy $\leq \pm 2\%$. When we combine the errors in
quadrature, we find that the typical error are of $20$\%. The systematic effects will not considered in this preliminary analysis. Systematic contributions
are: SZ calibration $\pm 8$\%, X-ray flux calibration $\pm 5$\%, radio halos $+3$\% and x-ray temperatute calibration $\pm 7.5$\%. For the sake consistency, the Hubble parameter
$H_{0}$  has been fixed by its best fit value $H^{*}_{0}=80
km/s/Mpc$.

\subsection{Linear Parameterization: $q = qo + q_{1}z$} 

In Figs. 1(a) and
1(b) we show, respectively,  the contour in the plane $q_{0}-q_{1}$
($68.3\%$, $90\%$ and $95.4\%$ c.l.) and likelihood of the
transition redshift from the Bonamente et al. sample. The confidence
region (1$\sigma$) are $-2.6 \leq q_0 \leq -0.25$ and $13 \leq q_1
\leq -3$. Such results favor a Universe with recent acceleration
($q_0 < 0$) and a previous decelerating stage ($dq/dz > 0$). From
Fig. 1(a) we see that he best fits to the free parameters are $q_{0}
= - 1.35, q_{1} = 4.2$ while for the transition redshift is
$z_{t}=0.32$ (see Fig. 1(b)).  Note the presence of a forbidden
region forming a trapezium. The horizontal line at the top is
defined by $q_1 = 0$, which leads to an infinite (positive or
negative) transition redshift. Note also that the segment at $45\%$
defines the infinite future ($z_t = -1$). In addition, one may
conclude that the vertical segment on the left closing the trapezium
is also unphysical since it is associated $z_t \leq -1$,  thereby
demonstrating that the hatched trapezium is actually a physically
forbidden region (for a similar analysis involving luminosity
distance see \cite{CL08}). For comparison we have also indicated in
Fig. 1(a) the transition redshifts  $z_t= 0.15$ corresponding to a
flat $\Lambda$CDM with $\Omega_{\Lambda} = 0.43$, as well as, $z_t=
0.9$  corresponding to $\Omega_{\Lambda}\simeq 0.8$.

\subsection{2nd Parameterization: $q = qo + q_{1}z/1 + z$}

 In Figures
2(a) and 2(b) we display the corresponding plots for the second
parameterization. The confidence region (1$\sigma$) is now defined
by: $-2.4 \leq q_0 \leq -0.5$ and $13.5 \leq q_1 \leq 0$. Such
results also favor a Universe with recent acceleration ($q_0 < 0$)
and a previous decelerating stage ($dq/dz > 0$). As indicated in
Fig. 2(a), the best fits to the free parameters are $q_{0} = - 1.43$
and $q_{1} = 6.18$ while for the transition redshift is a little
smaller $z_{t}=0.3$ (see Fig. 1(b)). It should be noticed the
presence of the forbidden region (trapezium) with a minor difference
in comparison with Fig. 1(a), namely, as an effect of the
parameterization, the horizontal line now is at the bottom. Note
also that a decelerating Universe today ($q_0>0$) is only marginally
compatible at $2\sigma$ of statistical confidence.
\begin{figure}[t]
\resizebox{.45\textwidth}{!}
{\includegraphics{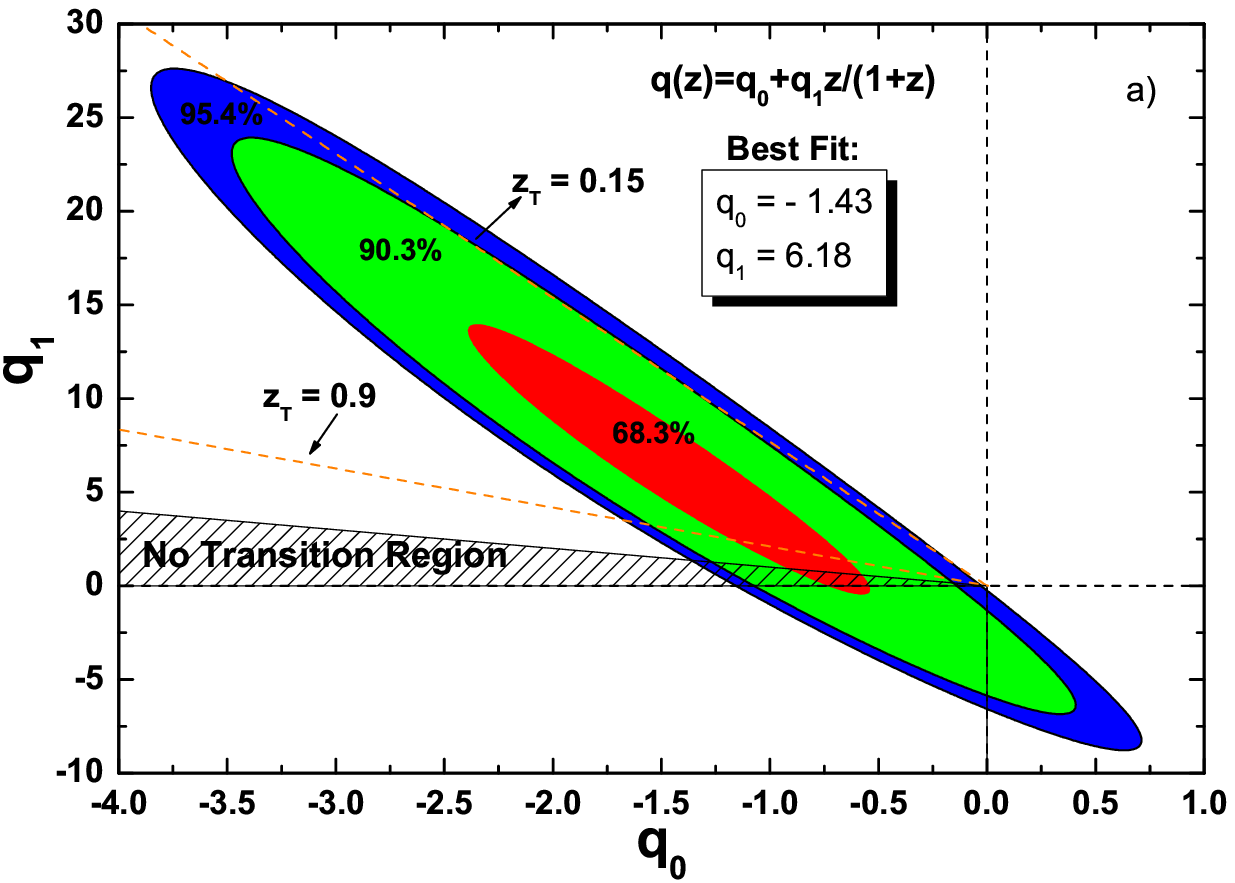}}
\resizebox{.45\textwidth}{!}
{\includegraphics{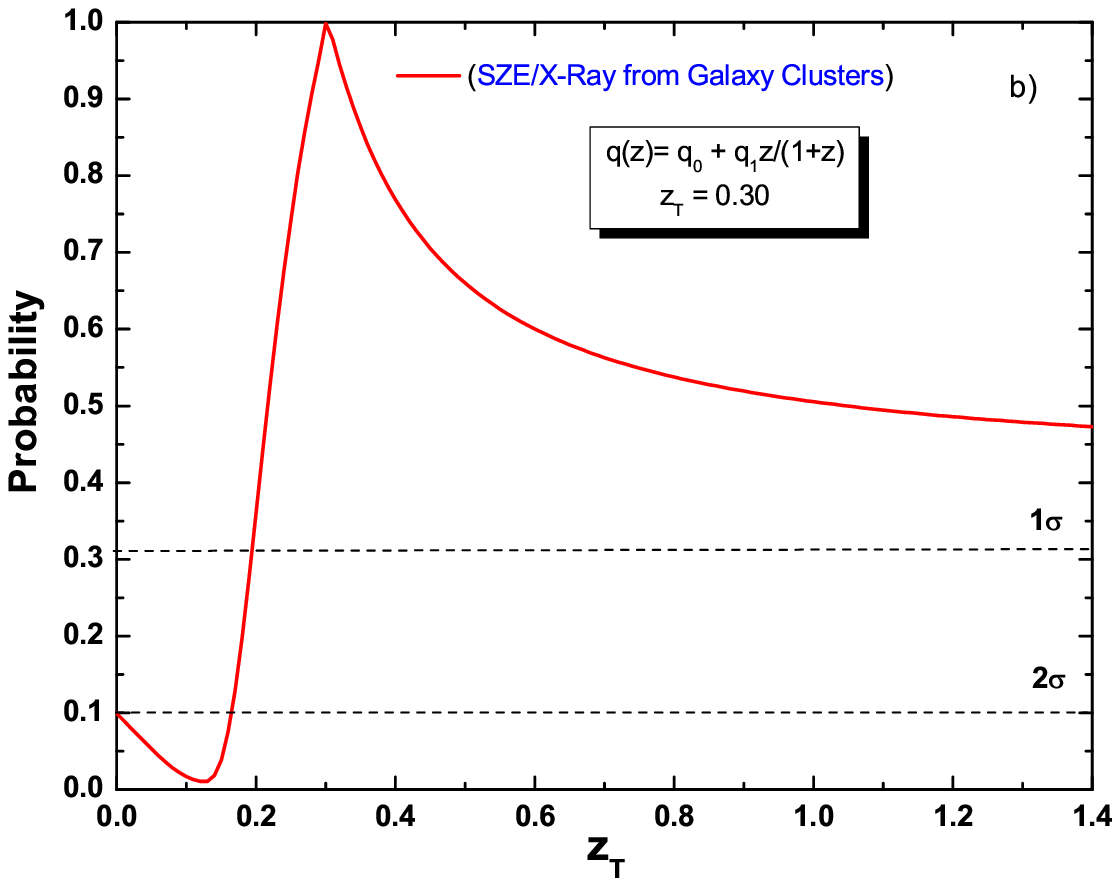}}
\caption{{\bf{a)}} Contours in the $q_0 - q_1$ plane for 38 galaxy
clusters data \cite{Boname06} considering $q(z)=q_0+q_1z/(1+z)$. The
best fit to the pair of free parameters is ($q_0,q_1$) =
($-1.43,6.18)$. {\bf{b)}} Probability for the transition
redshift and the associated best fit at $z_t=0.30$. Comparing with
Figs. 1(a) and 1(b)  we see that the results are weekly dependent on
the parameterizations.}\label{fig2}
\end{figure}

The results in the $q_{0}-q_{1}$ planes  for both cases suggest
that: (i) the Universe had  an earlier decelerating stage
($q_{1}=dq/dz
>0$ for linear case and $q_{1}>|q_{0}|$ for 2nd parameterization), and (ii) the Universe has been accelerating ($q_{0}<0$) since
$z\sim 0.3$. A similar result has been previously obtained using SNe
type Ia as standard candles by Shapiro and Turner\cite{TR02}.

\section{Prospects for Planck Satellite Mission} 

Let us now discuss
the potentiality of the SZE/X-ray technique when future data from
Planck satellite mission become available\cite{planck}. Planck
satellite mission is a project from European Space Agency whose
frequency channels were carefully chosen for measurements of thermal
Sunyaev-Zeldovich effect. In principle, the Planck satellite will
see (through SZE) about 30,000 galaxy clusters over the whole sky
with significant fraction of clusters near or beyond redshift unity.
However, since accurate angular diameter measurements require long
SZE/X-ray integrations, we do not expect that all observed clusters
might have useful distance measurements to constrain cosmological
parameters. Therefore, it is interesting to simulate two realistic
samples of angular diameter distances (ADD)  by using a fiducial
model to $D^{true}_{Ao,i}=D_{A}(z_{i}, q^{*}_{0}, q^{*}_{1},
H^{*}_{0})$, where $H^{*}_{0}$, $q^{*}_{0}$ and $q^{*}_{1}$ are the
best fit values to the linear case obtained from Bonamente {\it et
al.} sample\cite{Boname06}.

\begin{center}
\centerline{Table 1} \vspace{0.05cm}

  \begin{tabular}{c@{\quad}c@{\quad}c@{\quad}c@{\quad}c}
    $z$ range &  Clusters  &  bins  &  Clusters/bin  & $ADD_{i}$  Error \\
    & (P, O)& &(P,O) & (P,O) \\ \hline
    $[0.0, 0.5]$ & 100, 500 & 10 & 10, 50 & 15\%, 10\% \\
    $[0.5, 1.0]$ &  70, 350 & 10 & 7, 35    & 17\%, 12\% \\
    $[1.0, 1.5]$ &  40, 200 & 10 & 4, 20    & 20\%, 15\% \\

 \end{tabular}
\end{center}

The first simulation (termed pessimistic - P), assumes that only 210
clusters are distributed in the redshift ranges in the following
form: $0\leq z \leq 0.5$ (100), $0.5\leq z \leq 1$ (70) and $1\leq z
\leq 1.5$ (40) with ADD statistical errors of $15\%$, $17\%$ and
$20\%$, respectively (see Table 1).
In the second one (optimistic case - O), 1050 clusters were redshift
distributed as follows, $0\leq z \leq 0.5$ (500), $0.5\leq z \leq 1$
(350)  and $1\leq z \leq 1.5$ (200) with ADD statistical errors  of
$10\%$, $12\%$ and $15\%$, respectively. The redshift intervals were
partitioned into bins ($\Delta z=0.05$) with the clusters
distributed as shown in Table 1\cite{gol}. Both simulations were
carried out by marginalizing over the $H_{0}$ parameter in
$D_{A}(z_i,{\bf p})$ in Eqs. (1) and (5).

In Fig. 3 we display the results of our simulations for the linear
parameterization. The contours correspond $68\%$, $95\%$ and
$99.7\%$ c.l. for the optimistic (colored inner contours) and
pessimistic case (outer contours),  respectively. Comparing with
Fig. 1(a) we see that the allowed region was remarkably reduced even
in the pessimistic case. This means that ADD from SZE/X-ray will
become a potent tool for measuring cosmological parameters fully
independent and competitive with SNe type Ia.

\begin{figure}[t]
\resizebox{.55\textwidth}{!}
{\includegraphics{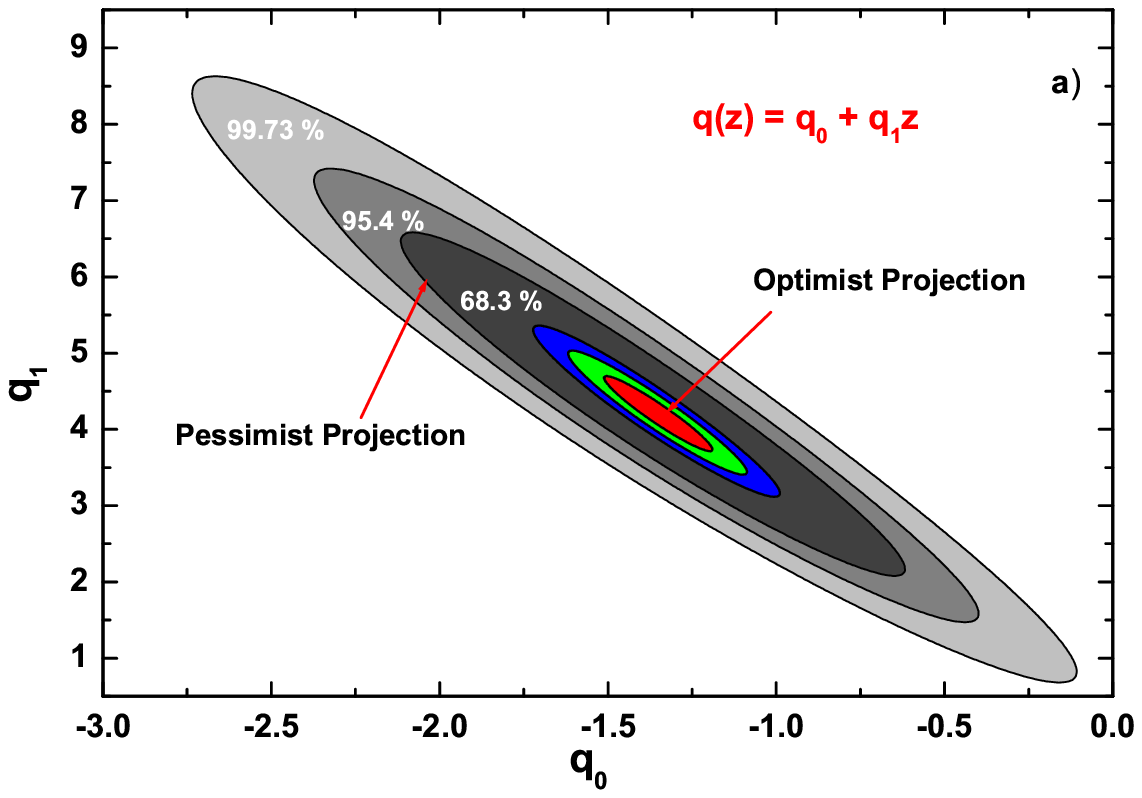}}
\resizebox{.52\textwidth}{!}
{\includegraphics{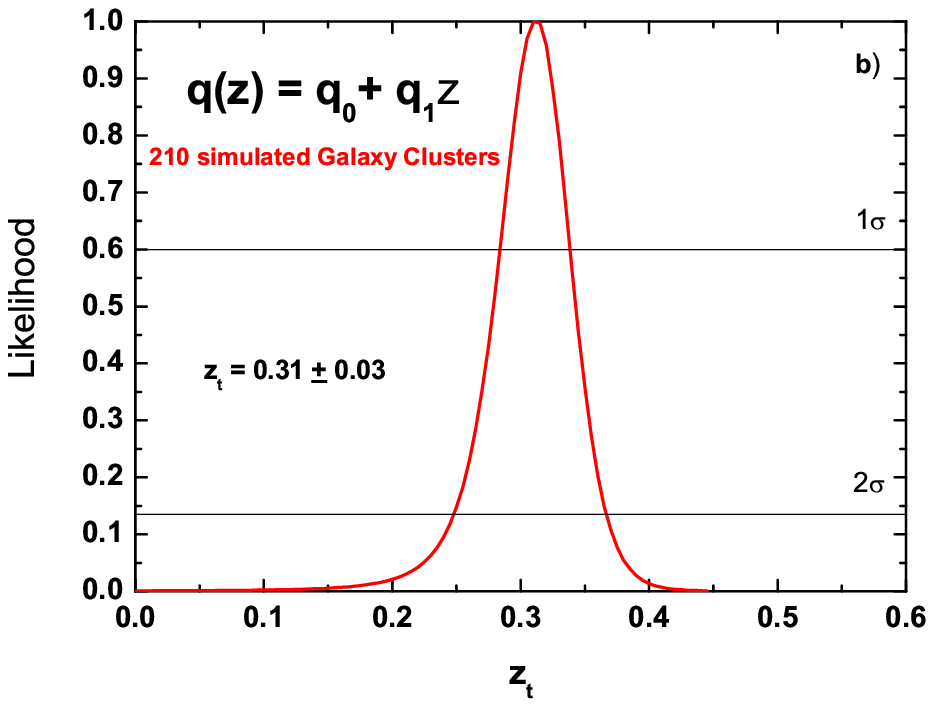}}
\caption{{\bf{a)}} we have contours on the plane
$(q_o,q_1)$ from synthetic Planck data as defined in Table 1. In the
pessimistic case, 210 P (pessimistic projection) and 1050 O
(optimistic projection) clusters were considered by assuming a
random error between 10\% and 20\% in the resulting angular diameter
distances.. {\bf{b)}}  we display the likelihood for the transition redshift in the pessimist projection. We obtain $z_{t}=0.31 \pm 0.03$.}\label{fig3}
\end{figure}
\section{Conclusions} 

We have shown that the combination of
Sunyaev-Zeldovich/X-ray data from galaxy clusters is an interesting
technique for accessing the present accelerating stage of the
Universe.  This result follows from a new kinematic approach based
on the angular diameter distance of galaxy clusters obtained from
SZE/X-ray measurements. By using two different parameterizations, it
was found the existence of a transition from a decelerating to an
accelerating expansion was relatively recent ($z_{t}\simeq 0.3$).

The ability of the future Planck satellite mission data to constrain
the accelerating stage was discussed by simulating two realistic
samples of angular diameters from clusters. The allowed regions in
space parameter was significantly constrained for both the
pessimistic and optimist simulations (Figure 3). The limits on the
transition redshift derived here reinforces the extreme interest  on
the observational search for the combination of Sunyaev-Zeldovich
and X-ray surface brightness from galaxy clusters.

Finally, it should also be stressed  that the present results
depends neither on the validity of general relativity nor the
matter-energy contents of the Universe and, perhaps, more important,
it is  also independent from SNe type Ia observations.

\begin{theacknowledgments}
JASL, RFLH and JVC are supported by FAPESP under grants, 04/13668-0,
07/5291-2, and 05/02809-5, respectively. JASL also thanks the
partial support by CNPq (Brazilian Research Agency).
\end{theacknowledgments}







\begin{thebibliography}{99}
\bibitem{riess98} A. G. Riess et al., \emph{Astron. J.} {\bf{116}},  1009-1038 (1998).
\bibitem{perl99} S. Perlmutter et al., \emph{Astrophys. J.} {\bf 517}, 565-586 (1999).
\bibitem{Padm03} T. Padmanabhan, \emph{Phys. Rep.} {\bf 380}, 235-320 (2003); P. J. E. Peebles and B. Ratra, \emph{Rev. Mod. Phys.}
{\bf{75}}, 559-606 (2003); J. A. S. Lima, \emph{Braz. J. Phys.} {\bf{34}}, 194-200
(2004); E. J. Copeland, M. Sami and S. Tsujikawa, \emph{Int. J. Mod. Phys.}
D {\bf 15},1753-1935  (2006); J. A. Frieman, M. S. Turner and D. Huterer,
\emph{Ann. Rev. Astron. \& Astrophys.} {\bf 46}, 385-432  (2008).
\bibitem{FR} L. Amendola, David Polarski and Shinji Tsujikawa,
\emph{Phys. Rev. Lett.} {\bf 98}, 131302-1 a 131302-4 (2007); T. P. Sotiriou and V.
Faraoni, \emph{Classical and Quantum Gravity} {\bf 25}, 205002-205018 (2008).
\bibitem{Kowalski08} M. Kowalski et al.,  \emph{Astrophys. J.} {\bf 686}, 749-778 (2008).
\bibitem{sunzel70} R. A. Sunyaev and Ya. B. Zel'dovich, \emph{Astrophys. Space Sci.} {\bf{7}}, 20-30 (1970);
R. A. Sunyaev and Ya. B. Zel'dovich, \emph{Comments Astrophys. Space Phys.}
{\bf{4}}, 173-178 (1972).
\bibitem{caval77} A. Cavaliere, L. Danese and G. De Zotti, \emph{Astrophys. J.}
{\bf{217}}, 6-15 (1977); A. Cavaliere and R. Fusco-Fermiano, \emph{Astron.
Astrophys.}, {\bf{70}}, 677-684 (1978); M. Birkinshaw,\emph{ Mon. Not. R.
Astron. Soc.} {\bf{187}}, 847-862 (1979).
\bibitem{Birk99} M. Birkinshaw, \emph{Phys. Rep.} {\bf 310}, 97-195 (1999).
\bibitem{SZEPapers2} J. G. Bartlett and J. Silk, \emph{Astrophys. J.} {\bf 423}, 12-18 (1994); J. E. Carlstrom, G. P. Holder and E. D. Reese, \emph{ARAA} {\bf 40}, 643-680 (2002); E. D. Reese  et al., \emph{Astrophys. J.} {\bf 581}, 53-85 (2002); M. E. Jones  et al., \emph{MNRAS} {\bf 357}, 518 (2002).
\bibitem{Boname06} M. Bonamente et al., \emph{Astrophys. J.} {\bf{647}}, 25-54 (2006).
\bibitem{CML07}J. V. Cunha, L. Marassi and J. A. S. Lima, \emph{MNRAS} {\bf 379}, L1-L5 (2007), [astro-ph/0611934].
\bibitem{TR02} M. S. Turner and A. G. Riess, \emph{Astrophys. J.}
{\bf{569}}, 18-22 (2002); C. Shapiro and M. S. Turner,  \emph{Astrophys. J.}
{\bf 649}, 563-569 (2006).
\bibitem{Riess04} A. G. Riess et al., \emph{Astrophys. J.} {\bf{607}}, 665-687 (2004).
\bibitem{EM06} Elgar{\o}y, {\O}., \& Multam{\"a}ki, T., \emph{JCAP}  {\bf 9}, 2-12 (2006) ;
A. G. C. Guimaraes, J. V. Cunha and J. A. S. Lima, \emph{JCAP} {\bf 10}, (2009) arXiv:0904.3550v1.
\bibitem{CL08} J. V. Cunha and J. A. S. Lima, \emph{MNRAS} {\bf 390}, 210-217 (2008), [arXiv:0805.1261].
\bibitem{Cunha09} L. Xu, C. Zhang, B. Chang and H. Liu, \emph{Mod. Phys. Let. A} {\bf 23}, 1939-1948 (2007); J. V. Cunha,  \emph{Phys. Rev. D} {\bf 79}, 047301
(2009).
\bibitem{Komat08} E. Komatsu et al., \emph{Astrophys. J. Suppl.} {\bf 180}, 330-376 (2009).
\bibitem{AbrSte72} M. Abramowitz and I. A. Stegun, {\it Handbook of Mathematical Functions}, Dover
Publications (1972).
\bibitem{planck}\emph{ The Scientific Programme of Planck}, the Planck Collaboration,  arXiv:astro-ph/0604069.
\bibitem{gol} M. Goliath, R. Amanullah, P. Astier, A. Goobar and R. Pain, \emph{Astron. Astrophys.} {\bf{380}}, 6-18  (2001).
\end{thebibliography}
\end{document}